\documentclass[aps,pra,reprint,superscriptaddress]{revtex4-1} %
\newif\ifarxiv 
\usepackage{natbib}
\usepackage{graphicx}
\usepackage{float}
\usepackage{amsfonts,amstext}
\usepackage{url}
\usepackage{xcolor}
\usepackage{braket}
\usepackage{soul}
\usepackage{amsthm}
\usepackage{physics}
\usepackage{hhline}
\usepackage{bbm}
\usepackage{enumerate}
\usepackage{comment}
\usepackage{colortbl}
\usepackage{amssymb}
\usepackage{comment}
\usepackage{float}
\usepackage{placeins}
\usepackage[colorlinks=true,linkcolor=blue,citecolor=blue,urlcolor=blue]{hyperref}
\usepackage{cleveref}

\newcommand{\notshortstack}[2][l]{%
  \begin{tabular}{@{}#1@{}}#2\end{tabular}%
}

\renewcommand{\emph}[1]{\textit{#1}}

\newtheoremstyle{theorem}
	{6pt}
	{}
	{\itshape}
	{}
	{\bfseries}
	{:}
	{.5em}
	{}
\theoremstyle{theorem}

\begin{document}

\title{Device Independent Quantum Key Activation}

\author{Bora Ulu}
\affiliation{Department of Applied Physics, University of Geneva, Switzerland}
\author{Nicolas Brunner}
\affiliation{Department of Applied Physics, University of Geneva, Switzerland}
\author{Mirjam Weilenmann}
\affiliation{Inria Saclay, Télécom Paris - LTCI, Institut Polytechnique de Paris, 91120 Palaiseau, France}
\affiliation{Department of Applied Physics, University of Geneva, Switzerland}

\date{\today}

\begin{abstract}
    Device-independent quantum key distribution (DIQKD) allows two distant parties to establish a secret key, based only on the observed Bell nonlocal distribution. It remains however, unclear what the minimal resources for enabling DIQKD are and how to maximize the key rate from a given distribution. In the present work, we consider a scenario where several copies of a given quantum distribution are jointly processed via a local and classical wiring operation. We find that, under few assumptions, it is possible to activate device-independent key. That is, starting from a distribution that is useless in a DIQKD protocol, we obtain a positive key rate by wiring several copies together. We coin this effect device-independent key activation. Our analysis focuses on the standard DIQKD protocol with one-way post-processing, and we resort to semi-definite programming techniques for computing lower bounds on the key rate.
\end{abstract}

\maketitle

The security of quantum cryptographic protocols can be demonstrated based on Bell nonlocality \cite{Ekert}. This leads to the possibility of ensuring secure key distribution assuming only relativistic causality \cite{bhk,acin2006}, as well as to device-independent quantum key distribution (DIQKD), where a secure key exchange is guaranteed without requiring a detailed description of the quantum devices \cite{Acin_2007,Pironio_2009,Arnon_Friedman_2019,Nadlinger_2022}.

In DIQKD, two distant users, Alice and Bob, perform local measurements on a shared entangled state, from which they can generate a secret key. The security of the key is verified via the observed Bell nonlocal correlations, which are used to upper bound the information of any adversary (Eve). A long-standing question is whether Bell nonlocality is enough to achieve DIQKD. Although initial proofs required a large degree of Bell inequality violation to guarantee security, recent works have shown that certain distributions with arbitrarily small Bell nonlocality can still enable DIQKD~\cite{Lewis_2024,Farkas2024}. Yet, there also exist nonlocal distributions that cannot be used for DIQKD (at least via the most commonly used protocol), as shown by the construction of an explicit attack~\cite{Farkas_2021,Lukanowski_2023}.

This raises two key questions. What are the minimal resources required for DIQKD, and how much secret key can be extracted from any given Bell nonlocal distribution.

\begin{figure}[t!]
    \centering
    \includegraphics[width=0.8\columnwidth]{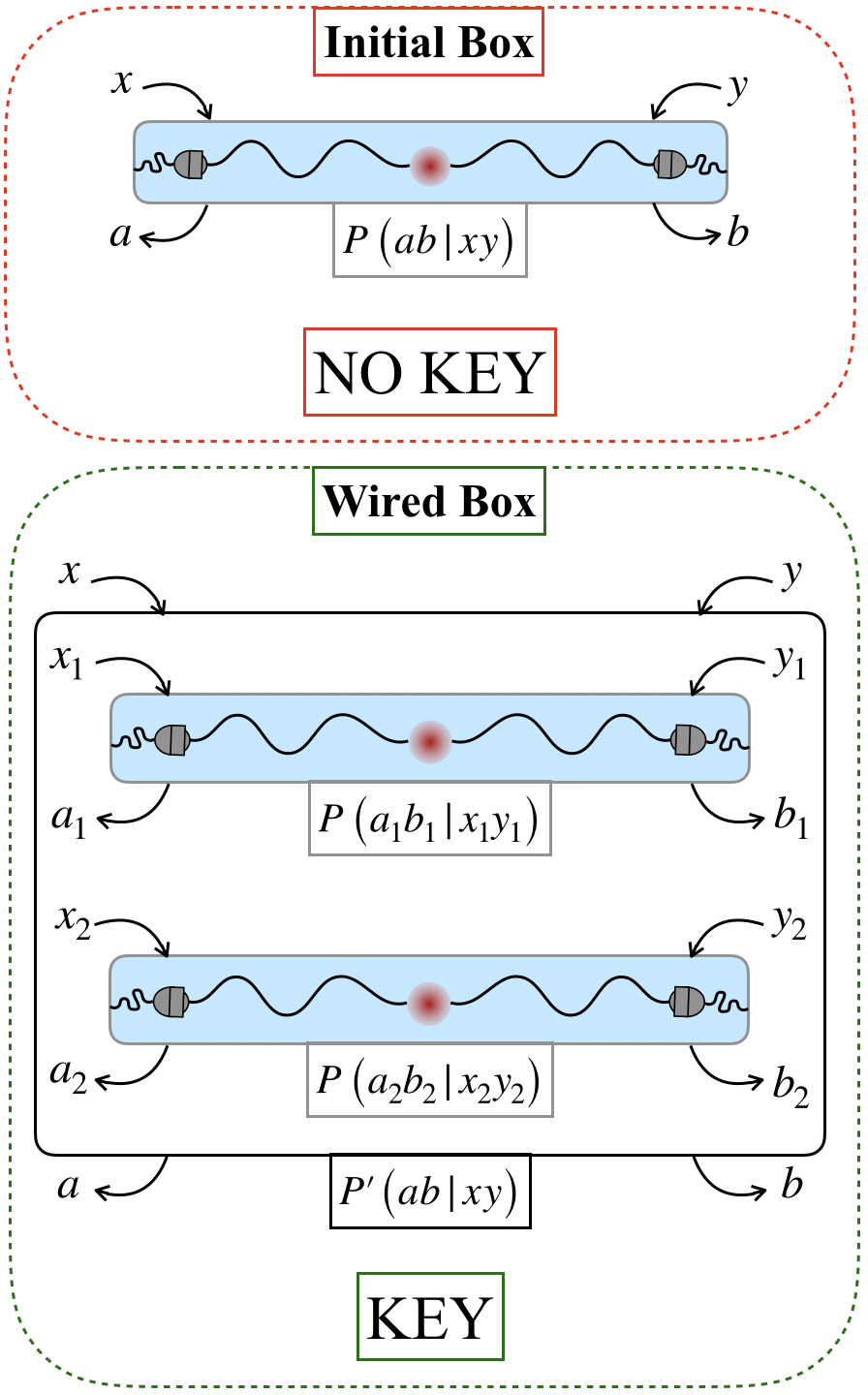}
    \caption{Consider an initial quantum nonlocal distribution $P(ab|xy)$ from which no DI key can be extracted. Here we ask whether by locally wiring several copies of $P(ab|xy)$, resulting in a different nonlocal distribution $P'(ab|xy)$, it becomes possible to implement DIQKD? We show that this is possible---under few assumptions in our analysis---leading to an effect of device-independent key activation. 
    }
    \label{fig1}
\end{figure}

In this work, we explore a novel approach for addressing these questions by considering a scenario where several copies of a given Bell nonlocal distribution are locally processed via classical wiring operations, as illustrated in Figure~\ref{fig1}. In particular, we investigate the possibility for activating DI key in the multi-copy scenario; that is, starting from a quantum distribution that cannot be used for DIQKD, can we nevertheless obtain a positive key rate by combining multiple copies of this distribution via local wirings? We present examples of this phenomenon---which we term device-independent key activation---under the following constraints: (i) we use the ``standard'' (most commonly used) DIQKD protocol \cite{Acin2006njp}, and (ii) we bound key rates using state-of-the-art semi-definite programming techniques \cite{Brown_2024}. Our approach is inspired by the protocol of nonlocality distillation, where the degree of Bell nonlocality of a distribution can be boosted via local wirings, see e.g.  \cite{Forster_2009,Brunner_2009,Allcock2009,Brito2019,Eftaxias_2023,Naik2023}. 
Additionally, we show that the power of specific attacks is reduced in the multi-copy regime, considering the examples of Ref.~\cite{Farkas_2021,Lukanowski_2023}. Overall, our work provides a fresh perspective on the problem of identifying the minimal resources necessary for DIKQD, and opens a number of questions that we discuss at the end of the paper.

\textit{Preliminaries.---} We consider the ``standard'' DIQKD protocol~\cite{Acin2006njp}. Alice and Bob share a bipartite entangled state $\rho$, and perform randomly chosen local measurements labeled by \textit{settings} $x\in\{0,1\}$ for Alice and $y\in\{0,1,2\}$ for Bob, 
with binary outcomes $a\in\{0,1\}$ and $b\in\{0,1\}$ respectively. 
These measurements are described by observables $A_{x}$ and $B_y$ with spectrum $\{\pm 1\}$, with associated projectors $M_{A}^{x,a} = \left[\mathbb{I}+(-1)^{a}A_x\right]/2$ for Alice and $M_{B}^{y,b} = \left[\mathbb{I}+(-1)^{b}B_y\right]/2$ for Bob, respectively. The resulting distribution is 
\begin{equation}\label{Eq: Born}
    P(ab|xy) = \text{Tr}\left[\rho \left(M_{A}^{x,a}\otimes M_{B}^{y,b}\right)\right].
\end{equation}
The raw key is generated via the pair of settings $(x=0,y=2)$, 
leading to an estimate of the quantum bit error rate given by $Q=P(a\neq b |x=0,y=2)$. To bound Eve's information on the key, Alice and Bob perform a Bell test for the settings $x\in\{0,1\}$ and $y\in\{0,1\}$, by estimating the Clauser-Horne-Shimony-Holt quantity for the distribution $P$: 
\begin{equation}
    S(P)= \left| \langle A_0 B_0 \rangle - \langle A_0 B_1 \rangle + \langle A_1 B_0 \rangle + \langle A_1 B_1 \rangle \right| ,
\end{equation}
where $\langle A_i B_j \rangle = \sum_{a,b} (-1)^{a \oplus b} P(ab|xy)$. 

Based on these quantities, Alice and Bob perform classical post-processing (via one-way communication from Alice to Bob) of the raw key. The final key rate can be lower bounded by the Devetak-Winter \cite{Devetak_2005} key rate $r \geq r_{DW}$ where
\begin{equation}\label{eq: DW}
    r_{DW}=H(A|E, X=0)-H(A|B,X=0,Y=2).
\end{equation}
The first term quantifies Eve's information about the key, which can be bounded from the CHSH value $S(P)$ \cite{Pironio_2009}. In this work, we consider a more general approach where the full distribution $P(ab|xy)$ is used, which gives tighter bounds in general. 
The second term quantifies the amount of error correction that Alice and Bob have to implement, which depends on the error $Q$. 


The performance of the above DIQKD protocol depends on the nonlocal properties of the distribution $P(a,b|x,y)$. The distribution resulting from appropriate Pauli measurements on a maximally entangled Bell pair, leading to maximal CHSH violation $S=2\sqrt{2}$, gives an optimal key rate of $r_{DW}=1$ \cite{Pironio_2009}. When the entangled state becomes noisy, then the key rate is reduced. More generally, the fact that $P(a,b|x,y)$ exhibits Bell nonlocality does not guarantee that DIQKD is possible, at least using the above protocol, as demonstrated via an explicit attack of Eve \cite{Farkas_2021}. 


\textit{Device Independent Key Activation.---} A central question is to characterize nonlocal distributions that enable DIQKD and understand how to use these distributions in the most effective way, i.e. optimizing the key rate. In this work, we uncover an effect that is relevant to both of these questions.

Consider a Bell nonlocal distribution $P(ab|xy)$ from which Alice and Bob cannot extract any secret key. In particular, using the above DIQKD protocol we get $r=0$. Now let us modify the scenario, and consider that Alice and Bob can process several (independent) copies of the same distribution $P(ab|xy)$ in each round of the protocol. Of course, the processing should be made locally by Alice and Bob, and should not involve any additional nonlocal resource such as classical communication. 

In this picture, it is convenient to represent each copy of the distribution $P(ab|xy)$ as a ``nonlocal box'', as sketched in Figure 1. Each nonlocal box takes inputs and outputs; for box $i$, we denote $x_i$ and $a_i$ the input and output of Alice, and similarly $y_i$ and $b_i$ for Bob.
The goal for each party is now to locally wire their nonlocal boxes, resulting in a final nonlocal box, i.e. a new distribution $P^{\prime}(ab|xy)$ which is in general different from the initial one $P(ab|xy)$. 

Our main goal is to show that the final nonlocal box $P^{\prime}(ab|xy)$ can be useful for DIQKD, i.e. lead to a strictly positive key rate, even though we started from a useless nonlocal box $P(ab|xy)$. In this way, the property of being useful in DIQKD can be activated by moving to the multi-copy regime. We coin this effect ``device-independent key activation''.

In the following we will present examples of device-independent key activation in the following context. First we consider the usual (one-way) DIQKD protocol discussed above. Second, we lower bound the key rate (i.e the Devetak-Winter quantity) using state-of-the-art semi-definite programming techniques~\cite{Brown_2024}. More precisely, we start by constructing quantum nonlocal boxes $P(ab|xy)$ and bound numerically their key rate; as this is a relaxation of the problem, we cannot guarantee in full generality that $r=0$ as we get only a lower bound on $r_{DW}$ (going to a higher level in the hierarchy may give a positive key rate). In turn, we construct an explicit local wiring protocol (for two and three copies of $P(ab|xy)$) and show that the key rate of the final box $P'(ab|xy)$ is strictly positive. Thus, 
we show that key activation is possible in practice: distributions $P(ab|xy)$ for which there are no known techniques to obtain a positive key rate can be wired into boxes $P'(ab|xy)$ for which there are.

Before moving to our example, let us understand what it means to wire several nonlocal boxes. Consider two copies of a box $P$ and label them $P_1(a_1b_1|x_1y_1)$ and $P_2(a_2b_2|x_2y_2)$ such that Alice has access to the classical bits $(a_1,a_2,x_1,x_2)$ and similarly Bob has $(b_1,b_2,y_1,y_2)$.
In an ordered wiring of two boxes, for each input $x$ Alice chooses an input for her first box $x_1$. She then chooses the input for her second box $x_2$ which may depend on $(a_1,x_1,x)$. Finally, based on  $(a_1,a_2,x_1,x_2,x)$ she chooses the final outcome $a$. Bob follows a similar procedure on his side and the parties obtain the new box $P^{\prime}(ab|xy)$ \footnote{Wirings are in general more complicated: the order in which Alice uses her boxes may depend on $x$ and similarly for Bob (for a formal treatment see also Appendix~\ref{app:wirings}). In wirings of more than two boxes, the ordering can further be dynamical, meaning that the ordering of boxes can depend on previous outcomes. Notice further that in the present work we are interested in initial and final boxes with the same input and output cardinality, thus restricting our considerations to such wirings.}. These local processings implemented by Alice and Bob can be represented as a map $\mathcal{F}$, which takes the two initial copies of $P$ to the final nonlocal box $P'$: $\mathcal{F}(P_1,P_2)=P'$. 
Note that the final box $P'$ has the same input/output cardinality as the initial box $P$, as we intend to use it for DIQKD via the above protocol. For a  mathematical characterisation of wirings and further details we refer to Appendix~\ref{app:wirings}.

\textit{Examples of key activation.---} We start by constructing a family of quantum distributions. Consider that Alice and Bob share the $4\times 4$-dimensional bipartite state
\begin{equation}\label{state}
    \rho_{\alpha, v} = \alpha \rho_v + \frac{1-\alpha}{2}\left(\ketbra{22}{22}+\ketbra{33}{33}\right),
\end{equation}
where $\rho_v= v \ketbra{\psi_0}{\psi_0} + (1-v) \frac{1}{2} \mathbb{I}$ with $\ket{\psi_0}=\frac{1}{\sqrt{2}} (\ket{01}-\ket{10})$. Hence the state consists of a noisy two-qubit Bell state (a Werner state $\rho_v$), mixed with a correlated noise term in an orthogonal subspace (spanned by $\{  \ket{2},\ket{3}\}$), with parameters $\alpha,v \in [0,1]$. The local measurements of Alice and Bob are given by the observables 
$   A_{x=0} = \sigma^{01}_{x}+\sigma^{23}_{z} \;,\; A_{x=1} = \sigma^{01}_{z} + \sigma^{23}_{z} \;,\; B_{y=0} = -\left(\sigma^{01}_{x}+\sigma^{01}_{z}\right)/\sqrt{2} + \sigma^{23}_{z}$ and $\; B_{y=1} = \left(\sigma^{01}_{x}-\sigma^{01}_{z}\right)/\sqrt{2} + \sigma^{23}_{z}$,
where the pair of indices $i,j \in\{0,1,2,3\}$ on the Pauli operator $\sigma^{ij}$ denotes the qubit subspace it acts on. Additionally, we set the third measurement of Bob to be $B_{y=2} = - \sigma^{01}_{x}+\sigma^{23}_{z}$, in order to minimize errors in the raw key.

From \eqref{Eq: Born} we get the resulting distribution, which it is convenient to express in the following form: 
\begin{align}\label{dist}
    P_{\alpha ,v}(ab|xy)=&\alpha v P_{T}(ab|xy)  + \alpha (1-v) P_0(ab|xy)  \\ \nonumber & + (1-\alpha) P_{C}(ab|xy) \, ,
\end{align}
where $P_{T}(ab|xy)$ denotes the distribution obtained in the noiseless case (i.e. setting $\alpha=v=1$ which leads to maximal CHSH value $S(P_T)=2\sqrt{2}$), $P_0(ab|xy)$ is the uniformly random distribution (white noise), and $P_{C}(ab|xy)$ is a distribution with perfectly correlated outputs, i.e. $P_C(00|xy)=P_C(11|xy)=1/2 $ for all $x,y$.

\begin{figure}[t]
    \centering
\includegraphics[width=0.95\columnwidth]{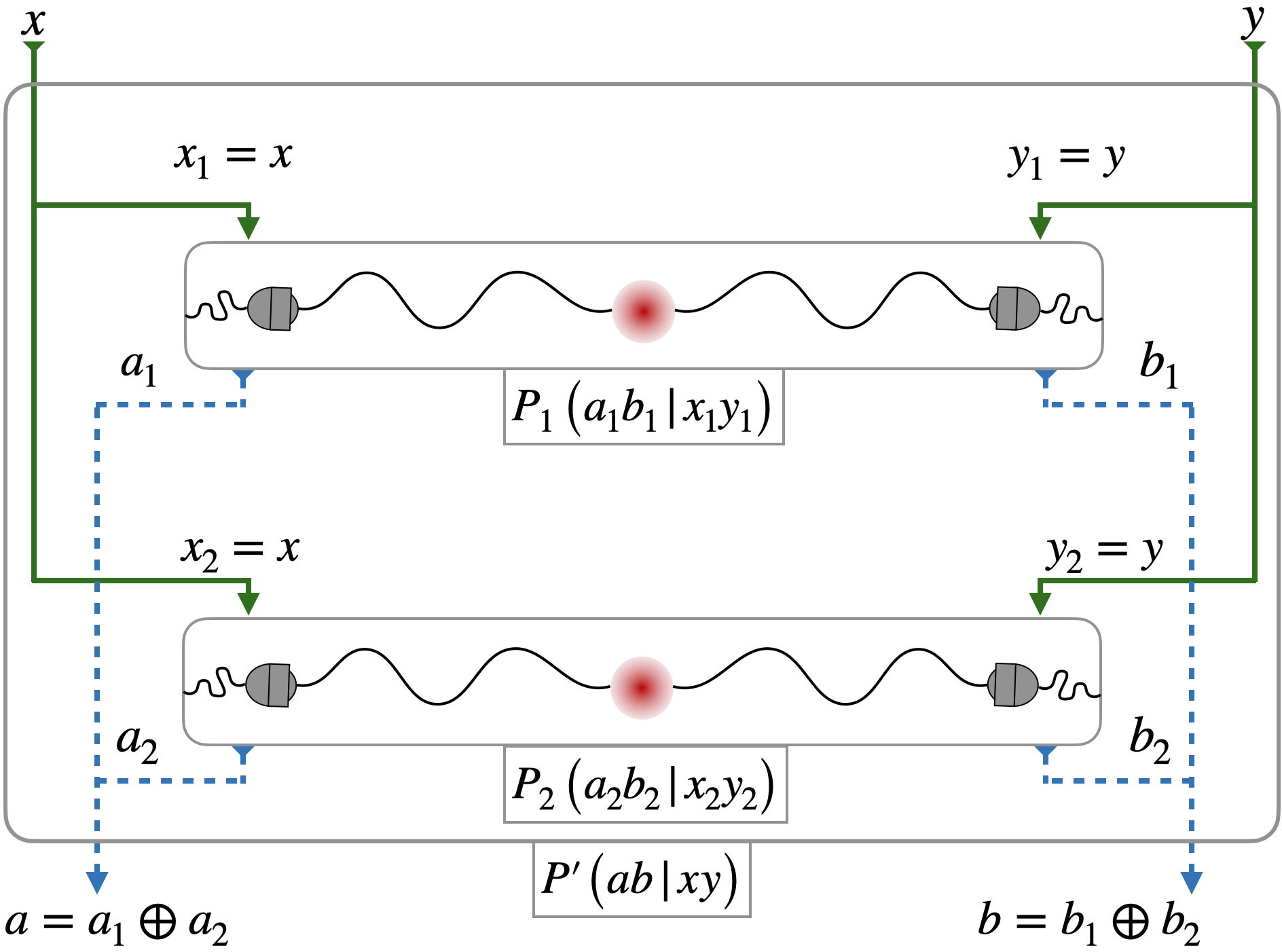}
    \caption{Sketch of the two-copy XOR-wiring~\cite{Forster_2009} (with flipped output bit). Alice forwards her input $x$ into both nonlocal boxes (green lines), and combines their outputs via an XOR operation, where the final output is $a=a_1 \oplus a_2$. Bob proceeds analogously. 
    }
    \label{wir}
\end{figure}

\begin{figure}[t]
    \centering
\includegraphics[width=0.95\columnwidth]{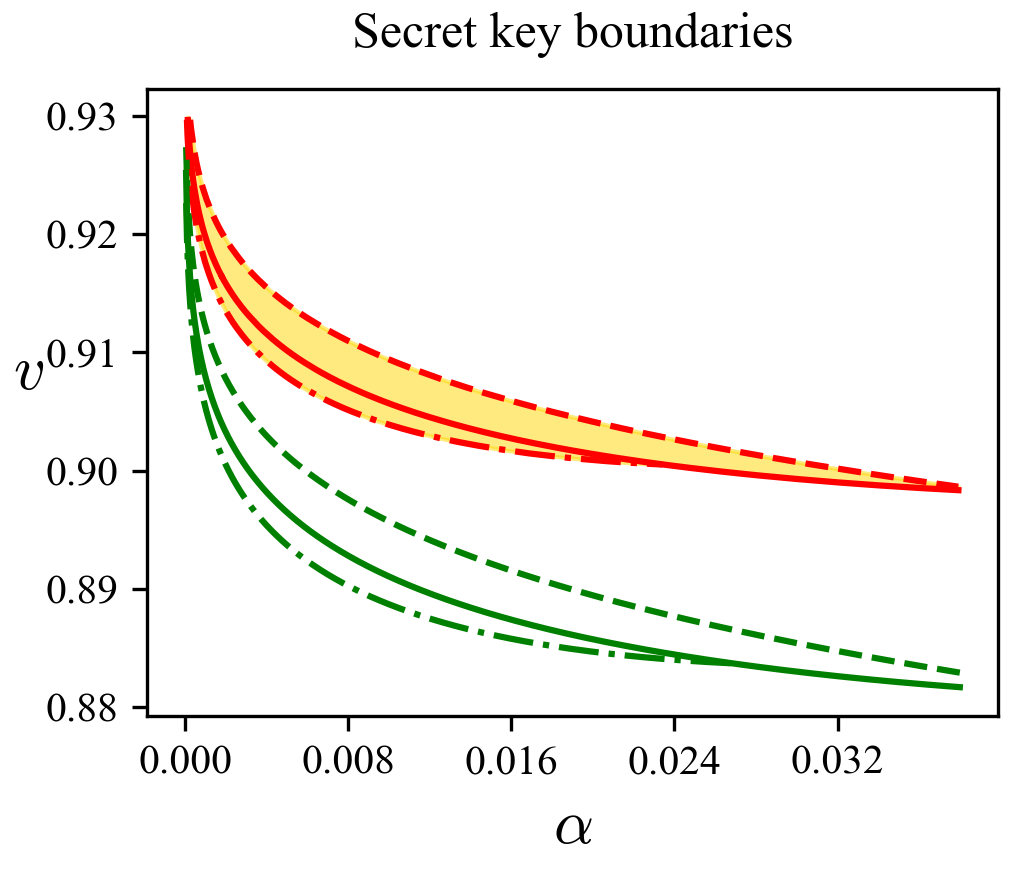}
    \caption{We consider nonlocal boxes as in Eq. \eqref{dist}, parametrized by $v$ and $\alpha$, and determine whether DIQKD is possible or not. In the single-copy regime, we get a positive key rate ($r>0$) above the red dashed curve. When wiring two (three) copies of the nonlocal box, we get $r>0$ above the red solid (dash-dotted) curve. Hence we see that DI key can be activated in the yellow region. The lower set of curves consider the case where Eve performs the specific attack in Ref. \cite{Farkas_2021}. In the single-copy regime, no key can be extracted ($r\leq 0$) below the green dashed curve. When wiring two (three) copies, we get $r\leq0$ below the green solid (dash-dotted) curve. Hence, certain nonlocal boxes that were initially vulnerable to the attack now become robust to it after wiring several copies.}
    \label{fig2}
\end{figure}

Now we compute the key rate \eqref{eq: DW} for the distributions $P_{\alpha, v}$. First, the error term is calculated analytically as
\begin{equation}\label{eq:8}
    H(A|B,Y=2,X=0) = h\left[1-\frac{\alpha(1-v)}{2}\right],
\end{equation}
where $h[p]=-p\log(p)-(1-p)\log(1-p)$ is the binary entropy. To estimate Eve's information, the first term in \eqref{eq: DW}, we resort to SDP techniques from Ref.~\cite{Brown_2024}, providing a lower bound on $H(A|E, X=0)$ from the full distribution $P_{\alpha, v}$ (see Appendix~\ref{app:lower_bounds} for more details). This allows us to get a lower bound on the key rate. We show in Figure~\ref{fig2} the parameter range, in terms of $\alpha$ and $v$, where $r>0$. Below the dashed red line, we find that they key rate is zero, suggesting that DIQKD is not possible with these distributions. 

\begin{figure}[t!]
    \centering
    \includegraphics[width=0.95\columnwidth]{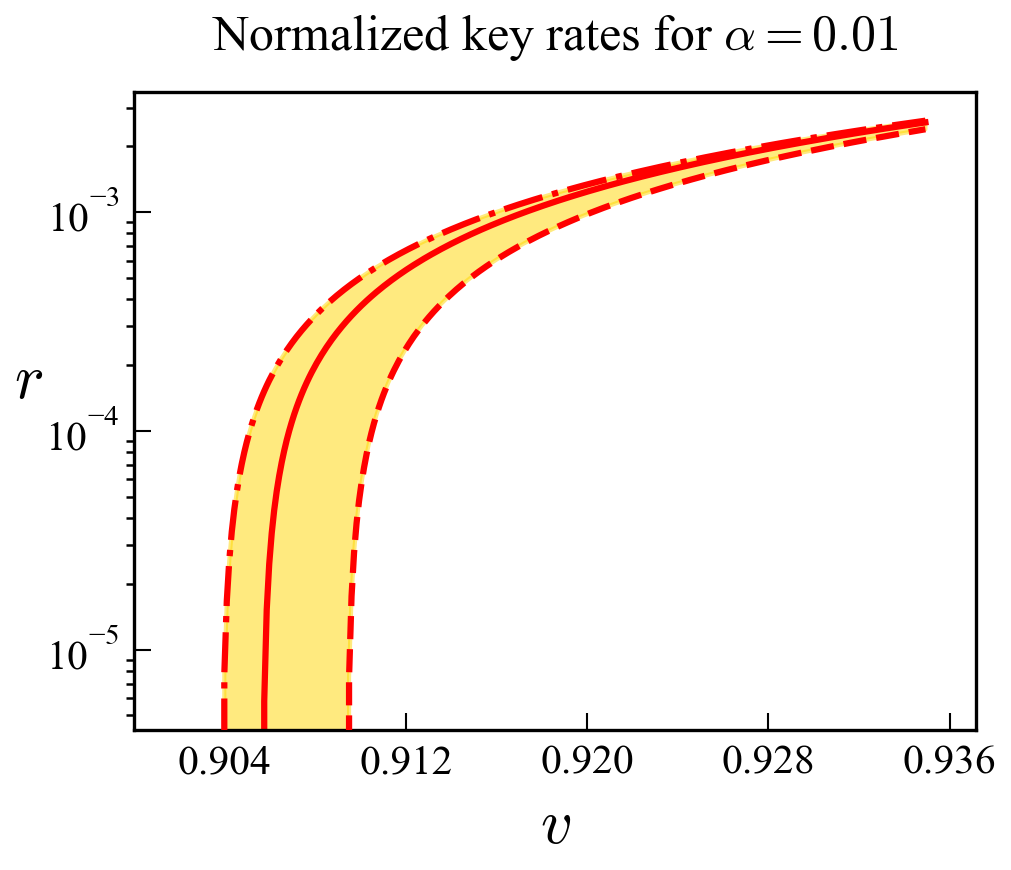}
    \caption{Lower bounds on the key rate $r$ (via SDP) as a function of the noise parameter $v$, for nonlocal boxes of the form in Eq.~\eqref{dist}; with $\alpha=0.01$. Wiring several copies of the nonlocal box allows for higher rates. Note that here $r$ is normalized by the number of copies used in each round. 
    }
    \label{fig3}
\end{figure}

Now let us move to a scenario where Alice and Bob process two copies of the distribution $P_{\alpha, v}(ab|xy)$. They use a so-called XOR local wiring \cite{Forster_2009}, described in Figure~\ref{fig3}. Through this process, they obtain a final distribution $P'(ab|xy)$, which turns out to also allow for a decomposition as in \eqref{dist}: the final distribution is $P'_{\alpha^{\prime},v^{\prime}}$ with parameters \begin{align}\label{W1}
     &\alpha^{\prime} = 1-(1-\alpha)^{2}-\frac{\alpha^{2}v^{2}}{2} \nonumber \\
     &v^{\prime} = \frac{2\alpha(1-\alpha)v}{\alpha^{\prime}}.
 \end{align}
All details can be found in Appendix-\ref{w_effect}. 

Next, we compute the key rates (following the same procedure as for the single copy case, see also Appendix~\ref{app:lower_bounds}) and determine the parameter range where key can be extracted. The results are shown in Figure~\ref{fig2}, showing clearly that the two-copy case (solid red line) increases the region where key can be distilled. Hence, we conclude that that there exists a range of distributions $P_{\alpha, v}$ (yellow region) which (i) give no key for the single copy case, but (ii) give a strictly positive key rate in the two-copy case. This is an example of DI key activation. 

It is also natural to investigate the case where more than two copies are processed in each round. We find that going to the three-copy regime (using again an XOR wiring) is slightly advantageous (red dash-dotted line in Figure~\ref{fig2}). However, going to four (or more) copies does not bring any further improvement; in this case, the noise term becomes too large.

A further question of interest is to see whether the key rate can be boosted via the wiring procedure. That is, starting from a distribution giving a low (but non-zero) key rate, can we significantly increase the performance. This is particularly relevant if the final key rate is more than $k$ times the initial one, when we wire $k$ copies. To investigate this question, we plot in Figure-\ref{fig3} the key rate $r$ as a function the noise parameter $v$, setting the other parameter to a fixed value $\alpha=0.01$. We observe that the key rate can be strongly boosted, even by several orders of magnitude, when the initial key rate is very low.

Finally, to get some intuition about the effect, it is insightful to compare DI key activation with nonlocality distillation. The goal of the latter is to increase the amount of nonlocality via wiring several copies, i.e. to get $S(P')>S(P)$. Here, the problem is different as we need to increase the key rate, in particular going from $r\leq 0$ to $r>0$. Wiring several copies may reduce Eve's information on the key (increasing the first term in (\ref{eq: DW})), but can also increase the error (the second term). It is nevertheless possible to get a positive balance, as we showed in the above example. However, due to the challenges in identifying distributions for which this is the case, we detail a heuristic procedure for finding such examples in Appendix~\ref{app:procedure_bora}.

 \textit{Fixed attack.---} A limitation of the above example is that we could not show analytically that the initial quantum distribution is useless for DIQKD; instead we had to resort to the best available numerical techniques. 

 To complement our analysis, we also investigated quantum distributions that are proven to be useless for DIQKD, via the demonstration of an explicit attack performed by Eve \cite{Farkas_2021,Lukanowski_2023}. These distributions turn out to be  of the form in equation (\ref{dist}). Hence we apply the same analysis as above; combining two (and three) copies of these distributions via the XOR wiring in Figure~\ref{fig3}, and evaluating the key rate following the analysis of Ref.~\cite{Farkas_2021,Lukanowski_2023}; see details in Appendix-\ref{ccat}. The results are shown in Figure~\ref{fig2} in the lower set of curves. We see that the power of Eve's attack is reduced when multiple copies are wired, in the sense that the region where the key rate is zero is reduced. This means that there exist quantum distributions that were initially vulnerable to the attack, which now become robust to this attack when wiring several copies together. This can be seen as DI key activation with respect to a fixed attack of Eve.

\textit{Conclusion.---} We have considered the question of whether device-independent key can be activated. Specifically, starting from a quantum distribution that is useless for DIQKD, can one nevertheless extract a secret key by locally wiring several copies of this distribution. 

We have shown examples of this effect. At this point, a limitation of our result is that we rely on semi-definite programming techniques to bound the key rate extractable from the initial quantum distribution (i.e. in the single-copy regime). In other words, according to today's state-of-the-art techniques, we cannot extract DI key from this distribution. Yet, it could still be the case that this distribution enables DIQKD in the single-copy regime. A main question for future research would be to prove DI key activation starting from a distribution that is provably useless. Another limitation of our analysis is that we focus on the standard DIQKD protocol with one-way classical post-processing. Considering more general protocols would also be interesting.


 A notable feature of our protocol is that it uses an ordered wiring. This means that Alice and Bob use their boxes in the same ordered sequence, one box after the other \footnote{In an ordered wiring, Alice, upon receiving input $x$ determines the input of the first box $x_1$, receives output $a_1$, and in turn computes the input to the second box $x_2$ and so on. Bob proceeds similarly, and uses the boxes in the same order as Alice.}. Hence such a protocol could in principle be easily implemented in practice, and combined with other techniques for boosting the key rate \cite{Ho_2020}. Notably, there exist more general classes of wirings (not ordered)~\cite{Giorgos_nonwirings}, and it would be interesting to see if they are useful for DI key activation, even though their implementation would be more complex.

Finally, another avenue to investigate is whether wirings could improve the performance of DI randomness generation protocols \cite{colbeck2009,Pironio2010}, as well as rates in the semi-device-independent setting.


\begin{acknowledgments}
We thank Peter Brown and Jean-Daniel Bancal for useful discussions. This work was supported by the Swiss National Science Foundation via the Ambizione PZ00P2\_208779 and the Swiss
State Secretariat for Education, Research and Innovation (SERI) under contract number UeM019-3. 
\end{acknowledgments}

\bibliographystyle{apsrev4-1}
\bibliography{main.bib}

\newpage
\onecolumngrid
\appendix

\section{Formal introduction to wirings} \label{app:wirings}

Formally, a wiring can be characterized by locally applied functions $\chi_x$ such that in general $\chi_x\left(a x_1 x_2 a_1 a_2\right)\in [0,1]$ and $\chi_x\left(a x_1 x_2 a_1 a_2\right)\in \{0,1\}$ for deterministic wirings. An example is given by the wiring $x_1=x_2=0$ and $a=a_1\oplus a_2$, which corresponds to $\chi_x\left(a x_1 x_2 a_1 a_2\right) = \delta_{x_1,0}\delta_{x_2,0}\delta_{a,a_1\oplus a_2}$. 
 
In general, Alice and Bob can use wirings characterized by $\{\chi_{x}(ax_1x_2a_1a_2)\}_x$ and $\{\xi_{y}(by_1y_2b_1b_2)\}_y$, respectively, to transform two copies of a probability distribution to another 
\begin{align}\nonumber
    P^{\prime}(a b | x y)=\sum_{\substack{x_1, x_2, y_1,y_2\\ a_1, a_2,b_1,b_2}} &P_{1}\left(a_1 b_1 | x_1 y_1\right) P_{2}\left(a_2 b_2  | x_2 y_2 \right)\\[1pt]\nonumber &\chi_x\left(a x_1 x_2 a_1 a_2\right) \xi_y\left(b y_1 y_2 b_1 b_2\right) 
\end{align}
For any wiring, the $\{\chi_{x}(ax_1x_2a_1a_2)\}_x$ have to satisfy the following conditions 
\begin{equation}\nonumber
0 \leq \sum_{\substack{a_1, a_2,\\ x_1, x_2}} \chi_x\left(a x_1 x_2 a_1 a_2\right) P\left(a_1 a_2 | x_1 x_2\right) \leq 1 \quad \forall x, a
\end{equation}
\begin{equation}
\sum_{\substack{a, a_1, a_2,\\x_1, x_2}} \chi_x\left(a x_1 x_2 a_1 a_2\right) P\left(a_1 a_2| x_1 x_2\right)=1 \quad \forall x 
\end{equation}
for all $P\in\{P_{i}^{\text{ex}}\}_{i}$, where $\{P_{i}^{\text{ex}}\}$ is the set of all extremal non-signalling boxes. This characterisation is tight in the 2-box 2-input 2-outcome case; for the case of more than two boxes it is not. Analogous conditions have to be satisfied by $\{\xi_y\left(b y_1 y_2 b_1 b_2\right) \}_y $.

The wiring from the main text (see Figure~\ref{fig3}) that allows us to demonstrate key activation is given by
\begin{equation}
        \chi_{x}(ax_{1}x_{2}a_{1}a_{2})= 
\begin{cases}
    1,& \text{if } \left(a_{1}\oplus a_{2} = a\right)\\ &\;\;\;\;\;\;\;\;\;\;\;\;\;\;\;\wedge \left( x_{0}=x_{1}=x\right) \\
    0,              & \text{otherwise}
\end{cases}
\label{w_alice}
\end{equation}
for Alice and the same wiring for Bob, where Bob applies this for all three settings $y\in \{0,1,2\}$.


\section{Bounding key rates for a specific example of 2-copy key activation}
\label{app:lower_bounds}
%
We use the recently developed method from~\cite{Brown_2024}. The authors construct an optimization problem with an objective that eventually converges to $H(A|E, X=0)$. They lower bound the entropy with $H(A|E, X=0) \geq H^{n}_{m}(A|E, X=0)$ by constructing the optimization problem
\begin{align}\label{eq:5}
    c_m + \sum_{i=0}^{m-1}\frac{w_{i}}{t_{i}\log(2)} \inf& \sum_{a=0}^{1} \text{Tr}\left[\rho\left(M_{A}^{0,a}\left(Z_{a,i}+Z_{a,i}^{*}+(1-t_i)Z_{a,i}Z_{a,i}^{*}\right)+ t_i Z_{a,i}Z_{a,i}^{*}\right)\right]\\\nonumber
    \text{s.t. }& \text{Tr}\left[\rho M_A^{a,x}M_B^{b,y}\right] = P(a,b|x,y) & \forall \; a,b,x,y\\\nonumber
    &\sum^{1}_{a=0}M_{A}^{x,a} = \sum^{1}_{b=0}M_{B}^{b,y}=\mathbb{I} & \forall \; x,y
    \\\nonumber
    &M_{A}^{x,a} \geq 0 \;\;\; M_{B}^{y,b} \geq 0& \forall \; a,b,x,y\\\nonumber &
    \left[M_{A}^{x,a},M_{B}^{y,b}\right] = \left[M_{A}^{x,a},Z_{a,i}\right] = \left[M_{B}^{y,b},Z_{b,i}\right]=0 & \forall \; a,b,x,y\\\nonumber
\end{align}
where $w_i$ and $t_i$ are Gauss-Radau quadratures with $m$ total nodes, $c_m = \sum_{i=1}^{m-1} \frac{w_{i}}{t_{i}\log(2)}$ and $Z_{a,i}$ are bounded operators. This is a non-commutative polynomial optimization problem, hence can be relaxed using the Navascués Pironio Acín (NPA) hierarchy~\cite{Navascu_s_2008}. There are two parameters in the relaxation, the number of nodes, which corresponds to the accuracy to which the logarithm in the von Neumann entropy is bounded, as well as the level of the NPA hierarchy. Hence, using more nodes as well as going to higher NPA levels makes the bounds from the relaxation tighter.

For fixed states and measurements, the objective of (\ref{eq:5}) converges to $H(A|E,X=0)$ if the infimum over $Z_{a,i}$ is taken outside the sum over the nodes as
\begin{equation}\label{hard_sdp}
    c_m + \inf_{Z_{a,i}} \sum_{i=0}^{m-1}\frac{w_{i}}{t_{i}\log(2)}  \sum_{a=0}^{1} \text{Tr}\left[\rho\left(M_{A}^{0,a}\left(Z_{a,i}+Z_{a,i}^{*}+(1-t_i)Z_{a,i}Z_{a,i}^{*}\right)+ t_i Z_{a,i}Z_{a,i}^{*}\right)\right].
\end{equation}
However, for the objective from \eqref{eq:5} a similar result is not known~\cite{Brown_2024}. 
Furthermore, 
relaxations of~\eqref{eq:5} via the NPA hierarchy may further reduce tightness of the bounds in the limit. This is because while the NPA hierarchy converges to the commuting operator model of quantum correlations~\cite{Navascu_s_2008}, in general this is not equal to the tensor product model~\cite{MIP_star}.

In comparison with using the min-entropy $H_{\text{min}}(A|E, X=0)=-\log\left(P_{g}(a|x=0,E)\right)$ where $P_{g}(a|x=0,E)$ is the eavesdropper's guessing probability to bound the von Neumann entropy $H(A|E, X=0)\geq H_{\text{min}}(A|E, X=x^{*})$, (\ref{eq:5}) gives superior bounds when a sufficient number of nodes ($m\geq 8$) are used. 

\bigskip 

Using these techniques, we obtain the  lower bounds on the von Neumann entropy $H^{n=2}_{m=12}(A|E, X=0)$ with parameters given by $\alpha=0.02$ and $v=0.90236$:
\begin{equation}\label{bw}
\begin{array}{ll}
\text{Before wiring:} & \text{After wiring:} \\
H^{n=2}_{m=12}(A|E, X=0)=0.0108 & H^{\prime n=2}_{m=12}(A|E, X=0)=0.0206 \\
H(A|B,Y=2,X=0)=0.0111 & H^{\prime}(A|B,Y=2,X=0)=0.0203 \\
r\geq -0.0003 & r^{\prime} \geq 0.0003
\end{array}
\end{equation}
Here the optimization for $H^{n=2}_{m=12}(A|E, X=0)$ was performed with the SDPA-GMP library for high precision computation \cite{inproceedings,Wittek_2015}. Each SDP in the optimization process was computed with a numerical precision up to 9 digits. This however, is not the only distribution in the cross section parametrized by $\alpha$ and $v$ where we see key activation (see Figure~\ref{fig2}). To plot the figures, the Mosek ~\cite{mosek} solver was used as it is much faster with sufficient numerical precision. See ~\cite{github} for the details of the code used in this work.

\section{Effect of the wirings from Figure-\ref{wir} of the main text on $P_{\alpha, v}$}
\label{w_effect}
 In this section we provide details on the effect of the wiring \eqref{w_alice} (see also Figure~\ref{fig3}) applied to two copies of $P_{\alpha, v}$, where we denote the effect of the wiring on boxes $P_1$, $P_2$ as $\mathcal{F}(P_1,P_2)$.
In the following we show that after applying these wirings the new distribution $P^{\prime}$ allows for the same parametrisation as $P_{\alpha, v}$ but with larger nonlocality. 
However, it's important to note that nonlocality distillation does not always lead to key activation. 

We can express the distribution $P_{\alpha, v}$ in terms of white noise $P_{0}=P_{v=0}$ and the PR-box

 \begin{equation}
    P_{PR}(ab|xy)=\left\{\begin{array}{ll}
\frac{1}{2} & \text { if } a \oplus b = (x\oplus 1)y \\
0 & \text { otherwise. }\end{array}\right.
\end{equation}
as $P_{\alpha, w} = \alpha(wP_{PR}+(1-w)P_{0})+(1-\alpha)P_C$ where $w=v/\sqrt{2}$. Therefore, starting with the two boxes $P_1=P_2=P_{a,w}$ the step-by-step effects of the given wirings are as follows

\begin{itemize}
  \item $\mathcal{F}(P_{PR},P_{PR})=P_C$. The first box outputs $a_1 \oplus b_1 = (x\oplus1)y$ while the second box outputs $a_2 \oplus b_2 = (x\oplus1)y$. The outputs after the wiring satisfy the relation $a\oplus b = a_1 \oplus b_1 \oplus a_2 \oplus b_2 = 0$.
  \item $\mathcal{F}(P_{PR},P_{C})=P_{PR}$, The first box outputs $a_1 \oplus b_1 = (x\oplus1)y$ while the second box outputs $a_2 \oplus b_2 = 0$. The outputs after the wiring satisfy the relation $a\oplus b = a_1 \oplus b_1 \oplus a_2 \oplus b_2 = (x\oplus1)y$. The same holds for $\mathcal{F}(P_{C},P_{PR})=P_{PR}$. Thus, when one box outputs correlated noise the wiring compensates for this noise by outputting the PR box.
  \item $\mathcal{F}(P_{PR},P_{0})=P_{0}$, The first box outputs $a_1 \oplus b_1 = (x\oplus1)y$ while the outputs of the second box are completely random. Since $a_2$ is random and uncorrelated with $b_2$ and vice versa, the outputs $a = a_1\oplus a_2$ and $b = b_1\oplus b_2$ are also uncorrelated and random. The same holds for $\mathcal{F}(P_{0},P_{PR})=P_{0}$, $\mathcal{F}(P_{C},P_{0})=P_{0}$, $\mathcal{F}(P_{0},P_{C})=P_{0}$ and $\mathcal{F}(P_{0},P_{0})=P_{0}$.
  \item $\mathcal{F}(P_{C},P_{C})=P_{C}$, Here we have $a_1 \oplus b_1 = 0$ and $a_2 \oplus b_2 = 0$ and thus $a\oplus b = a_1 \oplus b_1 \oplus a_2 \oplus b_2 = 0$.
\end{itemize}
 Using these steps we can construct the distribution after wiring two copies of $P_{\alpha, w}$ in terms of $P_{PR}$, $P_{C}$ and $P_{0}$. With $w=v/\sqrt{2}$ we can again express the resulting distribution in terms of the distributions $P_{T}$, $P_{C}$ and $P_0$. Thus, obtaining the expression in (\ref{W1}).

\section{Search algorithm for 2-copy key activation}
\label{app:procedure_bora}

 Nonlocality is the primary resource used for DIQKD protocols \cite{Zapatero_2023} therefore, to find examples of key activation we used distributions and wirings where the violation of the CHSH inequality can be distilled. It has been shown that nonlocality cannot be distilled along the isotropic line \cite{Beigi2015}, i.e., PR-boxes mixed with white noise. We have also noticed that in general wirings do not perform well in nonlocality distillation with distributions that are mixed with white noise therefore, we have further restricted our search to distributions closer to the boundaries of the quantum set. Given an initial distribution $P_{\text{in}}(ab|xy)$ we first find the CHSH inequality $S(P_\text{in})\leq2$ maximally violated by this distribution. We then find the optimal wirings to distill the nonlocality of this distribution via the optimization problem,

\begin{align}\nonumber
\text{maximize}_{\xi_{y},\chi_{x}}\quad  & S(P^{\prime}) &\\ \label{sum}
\text{subject to}\quad& P^{\prime}(a b | xy)=\sum_{\substack{ a_1, a_2,b_1,b_2\\ x_1,x_2, y_1,y_2}} P_{\text{in}}\left(a_1 b_1 | x_1 y_1\right) P_{\text{in}}\left(a_2 b_2  | x_2 y_2 \right) \chi_x\left(a x_1 x_2 | a_1 a_2\right) \xi_y\left(b y_1 y_2 | b_1 b_2\right)\\\nonumber
&0 \leq \sum_{\substack{b_1, b_2\\y_1,y_2}} \xi_y\left(b y_1 y_2  \mid b_1 b_2\right) P\left(b_1 b_2 \mid y_1 y_2\right) \leq 1 \quad \forall b\in \{0,1\},y\in \{0,1\}, P\in\{P_{i}^{\text{ex}}\}_{i}\\\nonumber
&0 \leq \sum_{\substack{a_1, a_2\\x_1,x_2}} \chi_x\left(a x_1 x_2  \mid a_1 a_2\right) P\left(a_1 a_2 \mid x_1 x_2\right) \leq 1 \quad \forall a\in \{0,1\},x\in \{0,1\}, P\in\{P_{i}^{\text{ex}}\}_{i}\\\nonumber
&\sum_{\substack{b_1, b_2, b,\\ y_1, y_2}} \xi_y\left(b y_1 y_2  \mid b_1 b_2\right) P\left(b_1 b_2 \mid y_1 y_2\right) = 1 \quad \forall P\in\{P_{i}^{\text{ex}}\}_{i}\\\nonumber
&\sum_{\substack{a_1, a_2, x\\ x_1, x_2}} \chi_x\left(a x_1 x_2  \mid a_1 a_2\right) P\left(a_1 a_2 \mid x_1 x_2\right) = 1 \quad \forall P\in\{P_{i}^{\text{ex}}\}_{i}\\\nonumber
\end{align}
Since this is a convex optimization problem the solution will be a set of extremal wirings $(\xi^{*}_{y},\chi^{*}_{x})$. In the 2-input 2-output scenario there are 32 sequential, 32 AND-gated, 8 XOR-gated, 8 one-sided and 2 constant extremal wirings making a total of 82 extremal values that $\chi^{*}_{x}$ can take for each $x\in\{0,1\}$. The same holds for Bob's wirings, therefore there are $82^{4}$ extremal pairs $(\xi_{y},\chi_{x})$. However, since the goal of this optimization task is to distill nonlocality in a non-trivial way we know that the solution will not be any of the constant wirings if we obtain an increase. Thus we have $80^{4}\approx 41 \times 10^{6}$ pairs of extremal wirings $(\xi_{y},\chi_{x})$ that could be solution $(\xi^{*}_{y},\chi^{*}_{x})$. 

\medskip

\begin{center}
\begin{tabular}{|c|c|c|}
\hline Wiring class &  Condition for $\chi_{x}\left(a, a_1, a_2, x_1, x_2\right)=1$ & Label of wiring for each $\tau, \sigma, \delta, \epsilon \in\{0,1\}$ \\
\hline Constant & $x_1=x_2=\mu, \quad a=\nu$ & $2\mu + \nu+1$ \\
\hline One-sided & $x_1=x_2=\mu, \quad a=a_{\nu+1} \oplus \sigma$ & $(4 \mu+2 \nu+\sigma+1)+4$ \\
\hline XOR-gated & $x_1=\mu, \quad x_2=\nu, \quad a=a_1 \oplus a_2 \oplus \sigma$ & $(4 \mu+2 \nu+\sigma+1)+12$ \\
\hline AND-gated & $
\begin{array}{l}
x_1=\mu, \quad x_2=\nu \\
a=\left(a_1 \oplus \sigma\right)\left(a_2 \oplus \delta\right) \oplus \epsilon
\end{array}$
 & $(16 \mu+8 \nu+4 \sigma+2 \delta+\epsilon+1)+20$ \\
\hline Sequential & $
\begin{array}{l}
x_{\mu+1}=\nu, \quad x_{(\mu \oplus 1)+1}=a_{\mu+1} \oplus \sigma \\
a=a_{(\mu \oplus 1)+1} \oplus \delta a_{\mu+1} \oplus \epsilon
\end{array}$
 & $(16 \mu+8 \nu+4 \sigma+2 \delta+\epsilon+1)+52$ \\
\hline
\end{tabular}
\end{center}

Since the product $\chi_x\left(a x_1 x_2 | a_1 a_2\right) \xi_y\left(b y_1 y_2 | b_1 b_2\right)$ is nonzero only for a small set of indices
\begin{equation}
    \Lambda := \left \{(x,a,a_1,a_2,x_1,x_2,y,b,b_1,b_1,y_1,y_2) | \chi_x\left(a x_1 x_2 | a_1 a_2\right) \xi_y\left(b y_1 y_2 | b_1 b_2\right)=1\right \},
\end{equation}
such that $|\Lambda|=64$, the optimization task can be solved efficiently using a search algorithm that only computes the non-zero terms in the sum (\ref{sum}) for all pairs of extremal wirings. The algorithm enumerates over the extremal wirings in a divide and conquer fashion, see ~\cite{github} for details. With this method it takes less than a second to find the optimal pair $(\xi^{*}_{y},\chi^{*}_{x})$ that maximizes the CHSH violation of the final box. 

Once we have maximally distilled nonlocality for an initial distribution $P_{\text{in}}(ab|xy)$ and obtained $P^{\prime}(ab|xy)$ we construct the set of necessary SDPs to optimize (\ref{eq:5}) and check whether we have an improvement in the security term. If so we move on to find the optimal wiring $\xi_2\left(b y_1 y_2  | b_1 b_2\right)$ for Bob's key generation rounds using the optimization task given in Appendix~\ref{app:error_term} below.

We follow these steps for a variety of chosen $P_{\text{in}}(ab|xy)$ until we find an example of key activation. In this process we sampled $P_{\text{in}}(ab|xy)$ from areas close to the edges of the quantum set as nonlocality distillation seems to be less effective as more white noise is added to $P_{\text{in}}(ab|xy)$. See ~\cite{github} for details of the implementation. 

\subsection{Calculation of the error term}
\label{app:error_term}

With the box $P(ab|xy)$, Alice uses her measurement setting $x=0$ and Bob uses $y=2$ in order to generate a shared key. Alice measures $A_{x=0}=\sigma^{01}_{x}+\sigma^{23}_{z}$ and given the state (\ref{state}) the optimal measurement to have maximally correlated outcomes for Bob is $B_{y=2} = -\sigma^{01}_{x}+\sigma^{23}_{z}$. For these measurement settings the parties obtain their outcomes according to the distribution
\begin{equation}\label{B1}
    P_{\text{in}}(ab|02) = (1-\alpha+\alpha v)P_{C}(ab|00)+\alpha(1-v)P_{0}(ab|00).
\end{equation}

Given that Alice performs the wiring (\ref{w_alice}) for $x=0$ optimizing Bob's wiring for maximal $P^{\prime}(a=b|02)$ is a linear optimization problem given by

\begin{equation*}
\begin{array}{ll@{}ll}
\text{maximize}_{\xi_{2}}  & P^{\prime}(a=b|02) &\\[3pt]
\text{subject to}& P^{\prime}(a b | 02)=\sum_{\substack{ a_1, a_2,b_1,\\ b_2, y_1,y_2}} P_{\text{in}}\left(a_1 b_1 | 0 y_1\right) P_{\text{in}}\left(a_2 b_2  | 0 y_2 \right) \chi_0\left(a 00 | a_1 a_2\right) \xi_2\left(b y_1 y_2 | b_1 b_2\right)\\[2pt]
&0 \leq \sum_{\substack{b_1, b_2\\ y_1, y_2}} \xi_2\left(b y_1 y_2  \mid b_1 b_2\right) P\left(b_1 b_2 \mid y_1 y_2\right) \leq 1 \quad \forall b\in \{0,1\}, P\in\{\Tilde{P}_{i}^{\text{ex}}\}_{i}\\
&\sum_{\substack{b_1, b_2, b\\ y_1, y_2}} \xi_2\left(b y_1 y_2  \mid b_1 b_2\right) P\left(b_1 b_2 \mid y_1 y_2\right) = 1 \quad \forall P\in\{\Tilde{P}_{i}^{\text{ex}}\}_{i}\\
\end{array}
\end{equation*}
where $\{\Tilde{P}_{i}^{\text{ex}}\}_{i}$ is the set of all extremal boxes in the 3-input 2-output scenario. Notice that in this case, like in the 2-input case, the constraints are tight. Furthermore, by convexity, the optimal wiring $\xi_{2}(by_{1}y_{2}|b_{1}b_{2})$ will be an extremal point of the polytope generated by the constraints. These extremal wirings in the 3-input 2-output case are given by the tabular below.

\begin{center}
\begin{tabular}{|c|c|c|}
\hline Wiring class &  Condition for $\xi_{2}\left(by_{1}y_{2}|b_{1}b_{2}\right)=1$ & \notshortstack{Label of wiring for each $\mu, \nu, \sigma, \delta, \epsilon \in\{0,1\}$ \\ and $\mu,\nu \in \{0,1,2\}$}\\
\hline Constant & $y_1=y_2=\mu, \quad b=\sigma$ & $2\mu+\sigma+1$ \\
\hline One-sided & $y_1=y_2=\mu, \quad b=b_{\tau+1} \oplus \sigma$ & $(4 \mu+2 \tau+\sigma+1)+4$ \\
\hline XOR-gated & $y_1=\mu, \quad y_2=\nu, \quad b=b_1 \oplus b_2 \oplus \sigma$ & $(6 \mu+2 \nu+\sigma+1)+16$ \\
\hline AND-gated & $
\begin{array}{l}
y_1=\mu, \quad y_2=\nu \\
b=\left(b_1 \oplus \sigma\right)\left(b_2 \oplus \delta\right) \oplus \epsilon
\end{array}$
 & $(24 \mu+8 \nu+4 \sigma+2 \delta+\epsilon+1)+34$ \\
\hline Sequential & $
\begin{array}{l}
y_{\tau+1}=\nu, \quad y_{(\tau \oplus 1)+1}=\left(b_{\tau+1} \oplus \sigma \right) + \mu \;(\text{mod } 3) \\
b=b_{(\tau \oplus 1)+1} \oplus \delta b_{\tau+1} \oplus \epsilon
\end{array}$
 & $(48 \mu+16 \nu+ 8\tau +4 \sigma+2 \delta+\epsilon+1)+106$ \\
\hline
\end{tabular}
\end{center}

When $\chi_0\left(a 00 | a_1 a_2\right)$ is set to be (\ref{w_alice}) and $P_\text{in}$ is set to be (\ref{dist}), the ideal wiring for Bob is given by the same XOR wiring
\begin{equation}\label{w_bob2}
        \xi_{2}(by_{1}y_{2}|b_{1}b_{2})= 
\begin{cases}
    1,& \text{if } \left(b_{1}\oplus b_{2} = b\right)\wedge \left( y_{0}=y_{1}=2\right) \\
    0,              & \text{otherwise}.
\end{cases}
\end{equation}
Since $\mathcal{F}(P_{C},P_{C})=P_{C}$ and $\mathcal{F}(P_{C},P_{0})=\mathcal{F}(P_{0},P_{C})=\mathcal{F}(P_{0},P_{0})=P_{0}$ this wiring together with Alice's wiring transforms the distribution (\ref{B1}) into
\begin{equation}
    P^{\prime}(ab|02) = (1-\alpha+\alpha v)^{2}P_{C}(ab|00)+\left(1-(1-\alpha+\alpha v)^{2}\right)P_{0}(ab|00)
\end{equation}

As for any distribution of the form $P(ab|02)=\beta P_C(ab|00)+(1-\beta)P_{0}(ab|00)$ the entropy is given by ${H(A|B,Y=2,X=0)=h\left[(\beta+1)/2\right]}$ we recover the error terms in the main text. Specifically, for the error term after wiring the boxes, we obtain
\begin{equation}
    H^{\prime}(A|B,Y=2,X=0) = h\left[\frac{\left(1-\alpha(1-v)\right)^2+1}{2}\right].
\end{equation}
 Since the error term before the wirings is given by (\ref{eq:8}) and since $\left(1-\alpha(1-v)\right)^2 < 1-\alpha(1-v)$, the error term always increases after these wirings. Therefore, in order to have any improvement in the key rate, the increase in the security term must be sufficiently large as to compensate for the increase in the error term, i.e.,
\begin{align}
   H^{\prime}(A&|E, X=0)-H(A|E, X=0)> \nonumber \\  &H^{\prime}(A|B, X=0, Y=2)-H(A|B,X=0, Y=2). 
\end{align}

 \section{ Key activation under a fixed attack}
\label{ccat}

The one way key rate $r$ can also be upper bounded by constructing an \textit{individual attack} for Eve. A recently developed, powerful individual attack used to upper bound the key rate is the convex combination attack \cite{Lukanowski_2023}. In a convex combination attack (CC) the eavesdropper provides the parties with a combination of local and non-local quantum boxes
\begin{equation}\label{cc_at}
    P(ab|xy) = (1-q_{L})P_{NL}(ab|xy)+q_{L}P_{L}(ab|xy),
\end{equation}
with the weight of the local box being $q_L$ with $0 \leq q_L \leq 1$. The parties observe $P(ab|xy)$ and remain oblivious to this construction. As the eavesdropper can construct the local part of the distribution as a convex combination of deterministic distributions, they can gain full information on the outputs of this part of the distribution while their information about the outcomes of the (non-local) quantum part of the distribution is restricted. Therefore, the eavesdropper chooses the boxes $P_{NL}$ and $P_L$ such that the weight of the local part $q_L$ is maximized. The key rate is then upper bounded by
\begin{equation}
    r\leq H_{\text{cc}}(A|E, X=0)-H(A|B,Y=2,X=0),
\end{equation}
where the entropy $H_{\text{cc}}(A|E, X=0)$ is fully determined by the marginals of the distribution $P(abe|xy)= (1-q_{L})P_{NL}(abe|xy)+q_{L}P_{L}(abe|xy)$.

If we restrict the eavesdropper to perform the CC attack, key activation can be shown analytically. If the parties share the box $P_{v=v^{*}}$ for some visibility $v^{*}$, maximizing the local part in a convex combination~\eqref{cc_at} gives the following mixture:
\begin{equation}
    P_{v=v^{*}}(ab|xy)=\frac{\sqrt{2}v^{*}-1}{\sqrt{2}-1}P_{v=1}(ab|xy) + \left(1-\frac{\sqrt{2}v^{*}-1}{\sqrt{2}-1}\right)P_{v=1/\sqrt{2}}(ab|xy),
\end{equation}
where $v=1/\sqrt{2}$ is the highest visibility for which the distribution remains local. Here we simply express $P_{v=v^{*}}$ in terms of the quantum distribution that maximally violates the CHSH inequality $P_{v=1}$ and the nearest local distribution $P_{v=1/\sqrt{2}}$. See \cite{Farkas2024} for a more detailed explanation regarding the optimality of this decomposition.  When the eavesdropper distributes $P_{v=1}(ab|xy)$ the key generation outcomes of Alice have the entropy $H_{v=1}(A|E) = 1$ and when $P_{v=1/\sqrt{2}}(ab|xy)$ is distributed Alice's key generation outcomes have the entropy $H_{v=1/\sqrt{2}}(A|E) = 0$. By construction of the CC attack $H_{v=v^{*}}(A|E)=(\sqrt{2}v-1)/(\sqrt{2}-1)$. In our case the box that the parties share has an additional correlated noise term $P_{\alpha, v}=\alpha P_{v} + (1-\alpha) P_{C}$. This extra correlated noise can be added to the local part of the mixture that the eavesdropper samples from as
\begin{equation}
    P_{\alpha, v}(ab|xy)=\alpha \left(\frac{\sqrt{2}v-1}{\sqrt{2}-1}\right)P_{v=1}(ab|xy) + \frac{\alpha(1-v)}{1-1/\sqrt(2)}P_{v=1/\sqrt{2}}(ab|xy)+(1-\alpha)P_{C}(ab|xy).
\end{equation}

Hence, the entropy of Alice's key generation outcomes given the eavesdropper becomes,
\begin{equation}
    H(A|E) = \alpha (\sqrt{2}v-1)/(\sqrt{2}-1).
\end{equation}

\twocolumngrid

\end{document}